\documentclass[a4paper,11pt]{article}
\pdfoutput=1 

\usepackage{jinstpub}
\usepackage{threeparttable}
\usepackage{booktabs}
\usepackage{multirow}
\usepackage{dcolumn}
\usepackage{arydshln}
\usepackage{upgreek}

\newcommand{\resulta}{$\lambda_t = \bigl(350{{}^{+60}_{~-0}}{}\,{\mathrm{(sys)}}\,
                     \pm\, 50\,{\mathrm{(stat)}}\bigr)\,\mathrm{\upmu m}$}

\title{Transmission of xenon scintillation light through PTFE}

\author[a,1]{D. Cichon,\note{corresponding author}}
\author[a,2]{G. Eurin, \note{now at CEA/Saclay, IRFU (Institut de Recherche sur
             les Lois Fondamentales de l'Univers), F-91191 Gif-sur-Yvette
             CEDEX, France}}
\author[a,1]{F. J\"org,}
\author[a]{T. Marrod\'an Undagoitia,}
\author[a]{N. Rupp}

\affiliation[a]{Max-Planck-Institut f\"ur Kernphysik, Saupfercheckweg 1,
                69117 Heidelberg, Germany}

\emailAdd{dominick.cichon@mpi-hd.mpg.de}
\emailAdd{florian.joerg@mpi-hd.mpg.de}

\abstract{Polytetrafluoroethylene (PTFE), also known as Teflon, is a common
          material used in the construction of liquid xenon detectors due to
          its high reflectivity for the VUV scintillation light of xenon.
          We present measurements of the transmittance of PTFE for xenon scintillation
          light with peak emission at a wavelength of 175 nm. PTFE discs of
          different thicknesses are installed in front of a photosensor in two
          setups. One is filled with gaseous xenon, the other with liquid
          xenon. The measurements performed with the gaseous xenon setup at
          room temperature yield a transmission coefficient of \resulta.
          This is found to be in agreement with the observations made using the
          liquid xenon setup.}


\begin{document}
    \setcounter{tocdepth}{1}
    \maketitle
    \flushbottom

    \section{\label{sec:intro}Introduction}
    In the search for rare processes, such as the elastic scattering of
    particle dark matter off nuclei~\cite{Undagoitia:2015gya} or the search for
    neutrinoless double beta decay~\cite{Pas:2015eia}, liquid xenon detectors
    have shown great performance and excellent
    sensitivity~\cite{Aprile:2018dbl,Albert:2017owj,Akerib:2017xx,Cui:2017xx}.
    These detectors reconstruct the energy deposited by particles using xenon
    scintillation signals. Such signals are typically detected with
    photosensors sensitive to vacuum ultraviolet (VUV) light.
    The scintillation wavelength of xenon is centered in the vacuum ultraviolet
    regime at $\mathrm{\big(174.8 \pm 0.1\,(stat.) \pm 0.1\,(syst.)\big)\,nm}$
    as pointed in recent spectroscopic measurements~\cite{Fujii_2015xx} 
    (in contrast to previous measurements that
    reported the peak emission closer to 178\,nm~\cite{Jortner:1965xsw, Basov:1970}).
    In order to efficiently collect the emitted photons in a detector, VUV
    reflectors are required. Current liquid xenon (LXe) detectors employ
    polytetrafluoroethylene (PTFE, also known as Teflon) for this
    purpose~\cite{Aprile:2017aty,Akerib:2019fml,Cao:2014jsa,Auger:2012gs} due
    to its excellent reflectivity properties. The reflectivity of PTFE is
    measured to be between 37\% and 100\% depending on the surface finishing
    and the surrounding
    medium~\cite{Silva:2009ip,Levy:2014,Akerib:2012ak,Neves:2016tcw}. While
    measurements in vacuum~\cite{Levy:2014} or argon
    atmosphere~\cite{Silva:2009ip} point to lower reflectivity values,
    the ones in LXe (direct measurements or extracted from MC/data comparisons)
    are close to 100\%~\cite{Levy:2014,Akerib:2012ak,Neves:2016tcw}.

    The reduction of experimental backgrounds is a major enterprise in
    experiments searching for rare processes. Radioactive trace impurities,
    such as uranium and thorium, are commonly found in most detector
    materials~\cite{Aprile:2017gs}. Their subsequent $\alpha$-decays can cause
    the production of radiogenic neutrons via $\mathrm{(\alpha, n)}$ reactions. They can
    pose a dangerous background for dark matter searches as they produce
    nuclear recoil signals identical to the ones expected by dark matter
    interactions~\cite{Undagoitia:2015gya}. Since the cross-section for
    $\mathrm{(\alpha, n)}$ reactions on fluorine is high~\cite{Heaton:1989:an}, the
    overall amount of PTFE needs to be minimized. This would potentially also
    reduce background from other radionuclides, such as $\gamma$-radiation
    originating from the decay of $^{214}$Bi, which is a limiting factor for
    neutrinoless double beta decay searches~\cite{Agostini:2020adk}. However, if a too thin sheet
    is employed, transmission of light through the PTFE could become
    significant, leading to signal losses. Furthermore, experiments using an
    outer veto detector, for example XENON100~\cite{Aprile:2011dd} or LZ
    (in construction)~\cite{Akerib:2019fml}, could suffer from light leaking
    into the veto region, making data analysis significantly more complex.
    We report here on a study of the transmittance of PTFE for xenon
    scintillation light.

    \section{Detector setups}\label{sec:setup}
    Two setups are employed to measure the transmittance of PTFE for xenon
    scintillation light. Most of the measurements were performed in a small
    chamber at room temperature filled with gaseous xenon (GXe). To check the
    behavior in LXe, additional tests were subsequently performed using a LXe
    time projection chamber (TPC).

\begin{figure}[h]
        \centering
        \includegraphics[angle=0,width=0.5\textwidth]{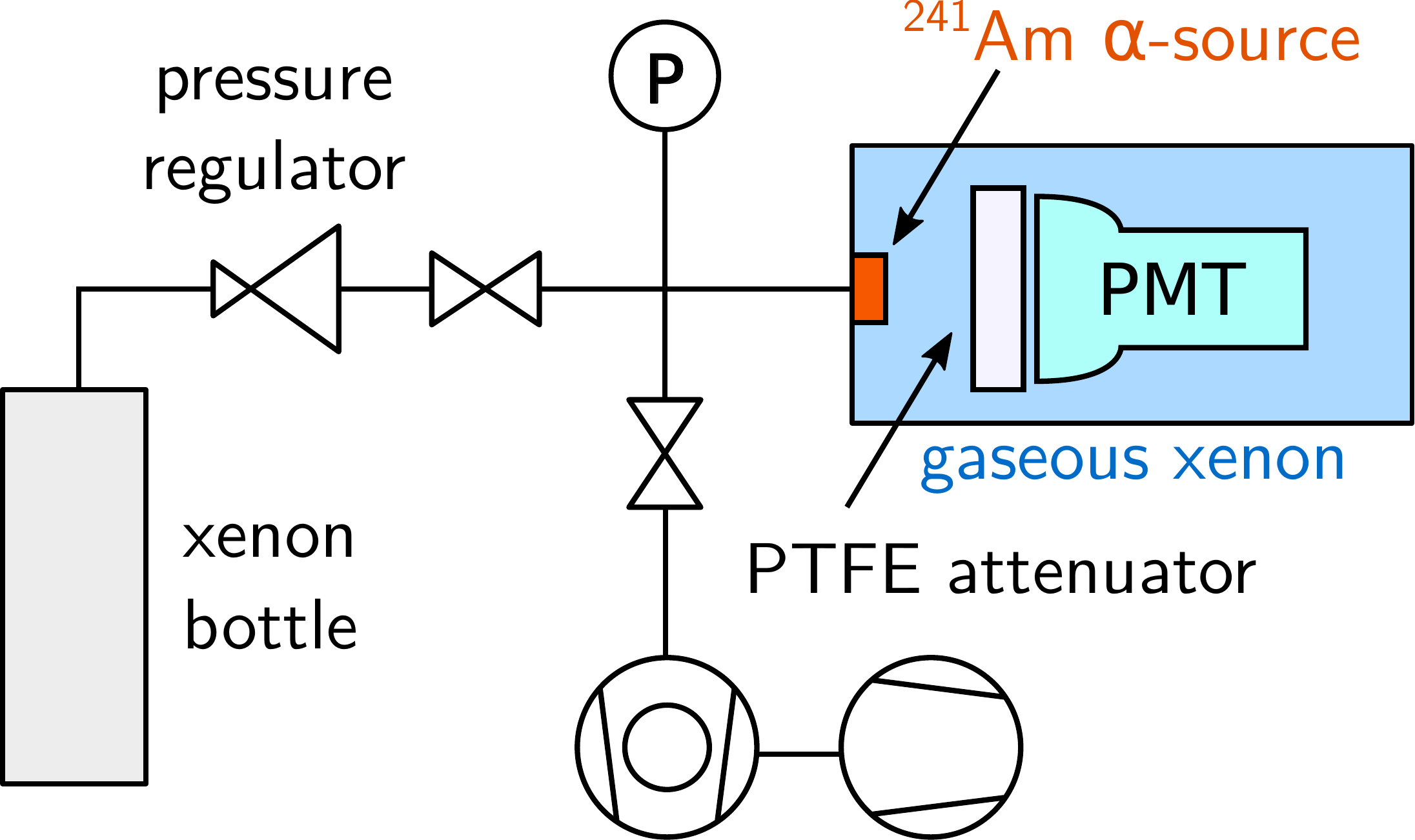}
        \caption[]{Schematic of the room
                   temperature setup.}\label{setup_gxe}
    \end{figure}

    Figure~\ref{setup_gxe} shows a schematic of the used room temperature
    setup.
    It consists of a 47 cm long stainless steel tube into which a structure
    holding both a photomultiplier tube (PMT) and a PTFE disc is inserted. The
    tube is filled with GXe at an average pressure of
    $\mathrm{(1098 \pm 7)\,\mathrm{mbar}}$. An $^{241\!}$Am $\alpha$-source
    ($Q_{\alpha} =$ 5.6\,MeV), mounted on the tube backplate, is used to
    produce scintillation light in the xenon gas. 
    Since the scintillation spectra of GXe and LXe are very similar~\cite{Jortner:1965xsw},
    this setup is expected to be a valid approximation.
    The distance between the $^{241\!}$Am source and the PTFE disc is fixed to
    $\mathrm{(14.0\,\pm\,0.3)\,cm}$ using three threaded rods. Scintillation
    photons are detected via a $\mathrm{3\,inch}$ Hamamatsu R11410-21 PMT~\cite{Barrow:2016doe}.
    Its signal is recorded via a CAEN V1724 digitizer triggered by an external
    discriminator. The PMT's gain has been measured at several voltages during
    the testing campaign described in~\cite{Barrow:2016doe}.

    PTFE discs in 6 different thicknesses, called attenuators from now on, were
    tested individually by placing them directly in front of the PMT window in
    order to measure the amount of scintillation light passing through. 
    All discs were manufactured from the same raw material 
    of extruded, natural PTFE~\cite{Auer:2020}, where half of them
    were made using a lathe with a backplate for supporting the material
    (batch 1). The other half was produced with a mill which utilized vacuum
    suction for material support (batch 2). The attenuator thicknesses were
    determined using a Mitutyo ID-F150 digital dial gauge. Each attenuator was
    measured at 49 points distributed uniformly in $\mathrm{(r^2,\varphi)}$
    over the area covering the PMT window. The results can be found in
    Table~\ref{tab:attenuators}. Attenuators belonging to the second batch
    show, on average, a more homogeneous thickness across their surface than the
    ones produced in the first batch. This can be explained by a better sample
    support due to the vacuum table during the production of batch 2. In
    addition, a measurement without attenuator was performed.

    \begin{table}[ht]
        \caption{List of the PTFE attenuators used in the GXe setup. Values
                 given for the thickness and its error are taken from the mean
                 and the standard deviation of the 49 measurements made for each
                 attenuator, respectively.}\label{tab:attenuators}
        \centering
        \begin{tabular}{c D{,}{\pm}{-1} c}
            attenuator	&	\multicolumn{1}{c}{thickness [$\mathrm{\upmu m}$]}  &	production batch\\
            \midrule
            1			&	97\,,\,2	&	2\\
            2			&	283\,,\,4		&	2\\
            3			&	442\,,\,6		&	1\\
            4			&	737.6\,,\,1.3	&	2\\
            5			&	965\,,\,4		&	1\\
            6			&	1310\,,\,30	&	1\\
        \end{tabular}
    \end{table}

    After each measurement, the setup needs to be opened to exchange the
    attenuator. Since impurities from the air, such as water, could cause
    additional attenuation of the scintillation light due to
    absorption~\cite{Ozone:2005ysa}, the opening is performed inside a
    gas-tight bag continuously flushed with nitrogen. To guarantee equivalent
    conditions among the different measurements, special attention was paid to
    the outgassing of the PTFE attenuators themselves. All of them were cleaned
    by applying a detergent and rinsing them three times using de-ionized
    water. Afterwards, the samples were submerged in ethanol and put into a
    drying chamber through which a continuous flow of 200\,SCCM nitrogen at a
    pressure of 50\,mbar is maintained. Exchange of each sample was done
    following a timed schedule, such that setup opening, evacuation, and xenon
    filling durations as well as the time between data taking and evacuation
    start were kept the same. The maximum deviation from this schedule during
    the measurement campaign is estimated to be smaller than 2 min.

\begin{figure}[h!]
        \centering
        \includegraphics[angle=0,width=0.7\textwidth]{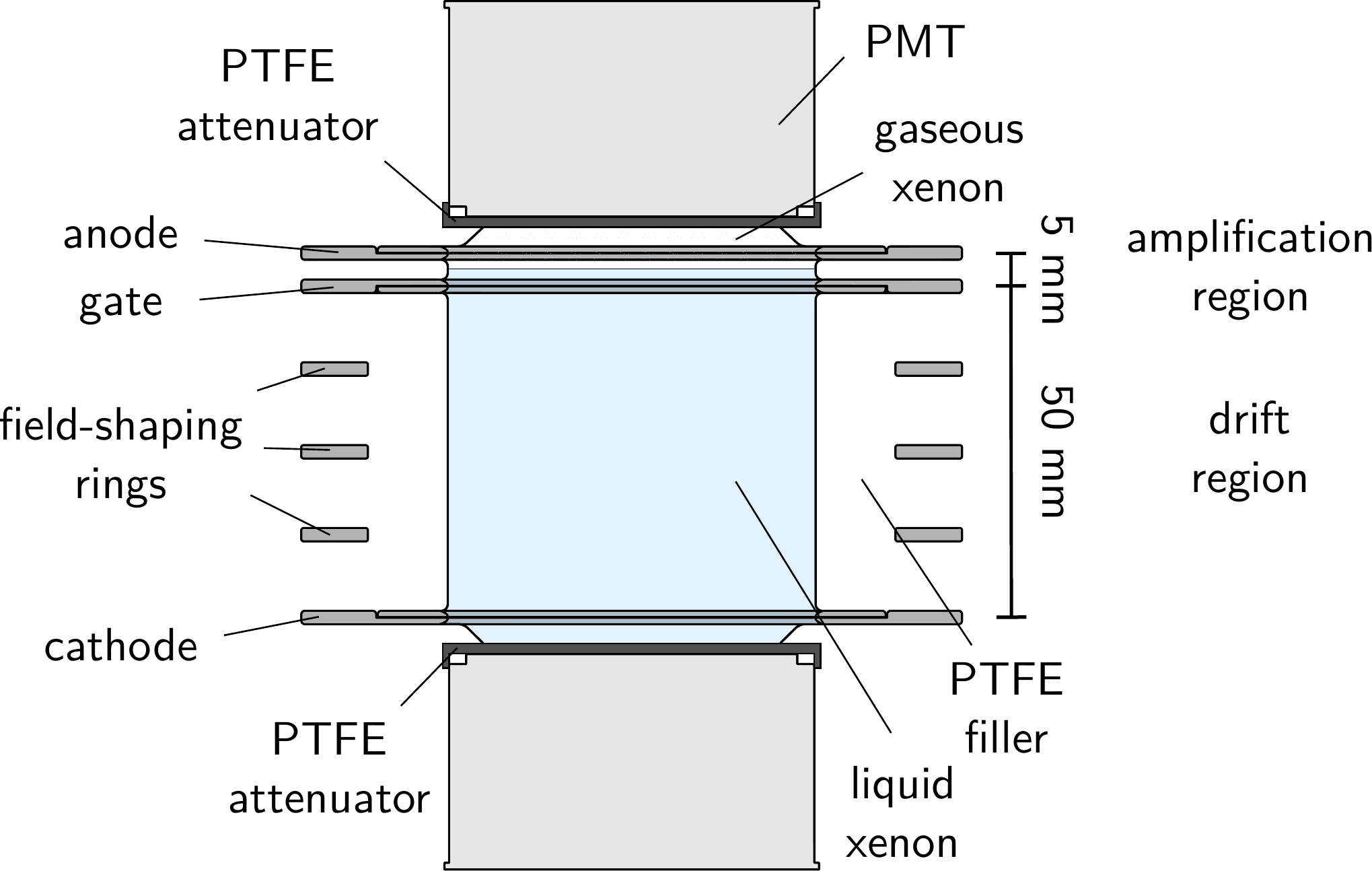}
        \caption[]{Simplified diagram of the HeXe TPC which is encased in
                   PTFE.\@ Attenuators are colored in black here for easier
                   visibility.}\label{setup_lxe}
    \end{figure}

    The results of the measurement described above are cross-checked with
    measurements using a local LXe TPC, Heidelberg Xenon
    (HeXe)~\cite{Cichon:2015,Joerg:2017}, illustrated in
    Figure~\ref{setup_lxe}.

    \noindent It contains a cylindrical sensitive volume of 5.6\,cm diameter
    and 5\,cm height enclosed by PTFE.\@ Electric fields are generated by three
    electrodes having a hexagonal meshed structure --- anode, gate, and
    cathode --- together with three field-shaping rings, all of them made out
    of stainless steel. Cathode, field-shaping rings and gate are connected in
    series via 1 G$\Upomega$ resistors and define a homogenous drift field by
    applying a voltage between cathode and gate, while a voltage applied
    between gate and anode creates the extraction field. To detect the
    scintillation light, two $\mathrm{2\,inch}$ Hamamatsu R6041-406 PMTs 
    are placed at the top and
    bottom of the sensitive volume, respectively. They are periodically
    calibrated \textit{in situ} following the technique described
    in~\cite{Saldanha:2016mkn}.

    For the measurements presented here, data from $^{83m}$Kr and $^{222}$Rn
    decays inside the TPC were acquired. $^{222}$Rn is extracted from an
    aqueous $^{226}$Ra solution by freezing it into an active charcoal trap,
    while $^{83m}$Kr is continuously emanated from $^{83}$Rb-containing zeolite
    beads~\cite{Manalaysay:2009yq,Kastens:2009rt} inside a stainless steel
    tube. Both unstable nuclides can individually be mixed into the xenon via a
    gas purification system, by directing the gas flow through either of the
    two sources. $\alpha$-decays from the $^{222}$Rn decay chain cannot be
    accurately measured in HeXe due to their scintillation signals saturating
    the PMTs. 
    To prevent this saturation, light attenuators made from PTFE were installed.
    The measurements in gaseous xenon at room temperature presented here were 
    carried out to find the PTFE attenuator thickness best suited for measuring 
    $\alpha$-decays with HeXe in liquid xenon.

    The attenuators were wiped with ethanol in an N\textsubscript{2}
    atmosphere and inserted into the TPC following the same opening procedure as in~\cite{Bruenner:2020}.

    \section{Data analysis}
    \subsection{Room temperature measurements}\label{subsec:room_temp_setup_data}
\begin{figure}[ht]
        \centering
        \includegraphics[trim=0cm 0cm 1.25cm 0cm, clip=true,angle=0,width=0.75\textwidth]{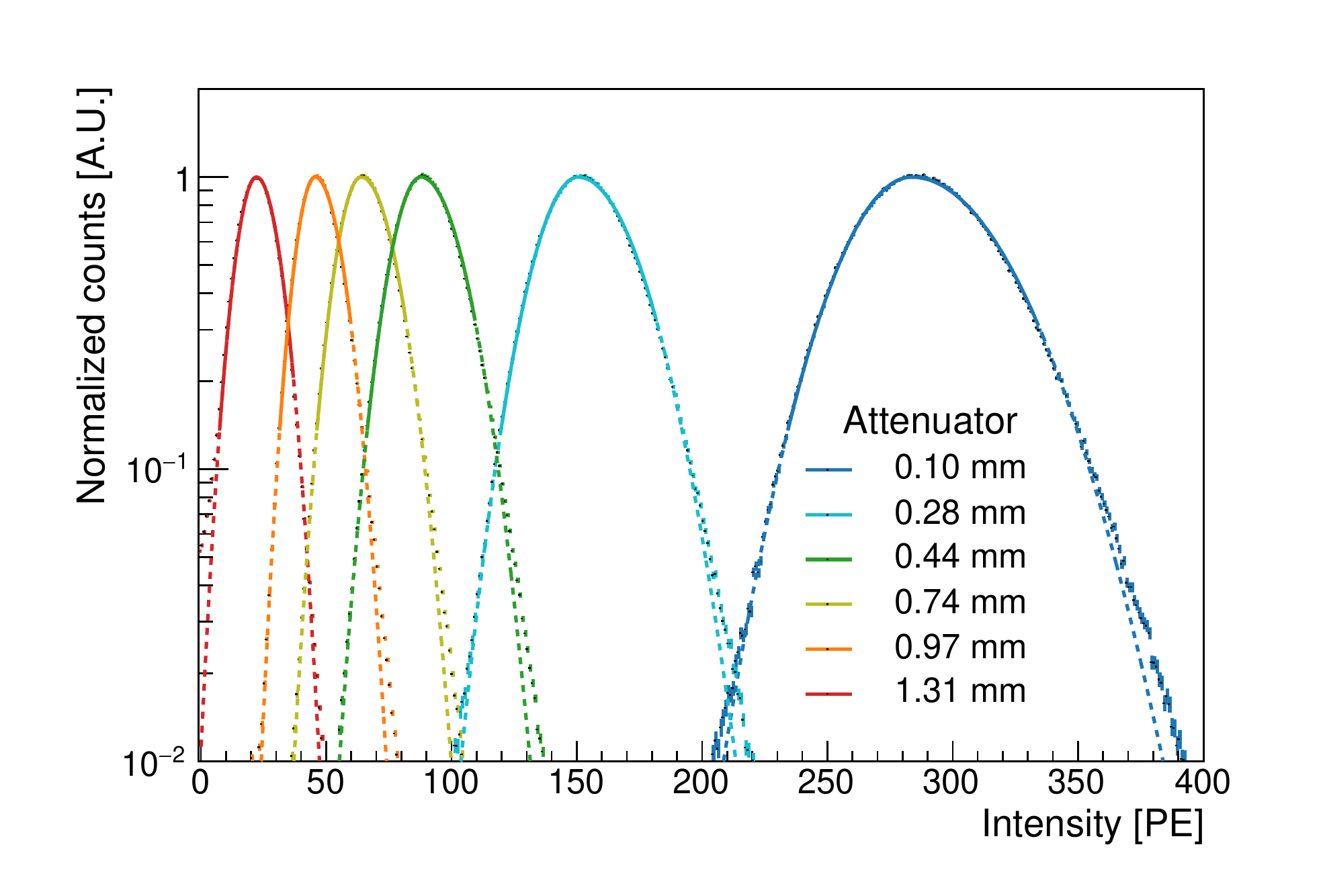}
        \caption[]{Full absorption peaks of $^{241\!}$Am $\alpha$-decays
                   measured in the room temperature setup. Spectra were
                   normalized to their respective peak height. A two-sided
                   Gaussian function is fitted to each peak, with the fit range
                   indicated by the solid lines.}\label{fig:full_abs_peaks}
    \end{figure}

    In the room temperature setup, the amount of scintillation light
    transmitted through the PTFE discs is determined by finding and integrating
    over the PMT signals using the same data processor as described
    in~\cite{Cichon:2015}.
    Figure~\ref{fig:full_abs_peaks} shows the part of the scintillation
    spectrum corresponding to the full absorption of the $^{241\!}$Am
    $\alpha$-particle for each attenuator (normalized to the peak maximum).
    The peak's shape is found to be well described by a two-sided Gaussian
    function of the form:

    \begin{align}
        f(x) = N\cdot
        \begin{cases}
            e^{-\frac{1}{2}\left(\frac{x-\mu}{\sigma_1}\right)^2} \qquad x<\mu\\
            e^{-\frac{1}{2}\left(\frac{x-\mu}{\sigma_2}\right)^2} \qquad x\geq\mu
        \end{cases}
        \label{eq:two-sided_gaus}
    \end{align}

    Here, $\mu$ represents the central peak position, with the $\sigma_{i}$
    representing the peak widths left and right to the central position
    respectively and $N$ being a scale factor. After a preliminary manual fit
    for extracting initial values for the widths ($\sigma_{i}'$), the range of
    the final fit is constrained to
    $[\mu - 2 \sigma_{1}', \mu + 2 \sigma_{2}']$ as indicated by the solid
    lines in Figure~\ref{fig:full_abs_peaks}. Dashed lines show the continuation
    of the functions outside of the fitted range and indicate reasonable agreement
    of the data with the fit. In all measured spectra, the peak is well above
    the cutoff caused by the threshold of the external discriminator used for
    triggering the acquisition of data.

    In addition to the full absorption peaks shown in
    Figure~\ref{fig:full_abs_peaks}, each spectrum exhibits two additional
    features. At twice the average signal size of each full absorption peak,
    an additional peak is observed which corresponds to two $\alpha$-decay
    signals from pile-up.
    This has been confirmed on signal waveforms. Another peak,
    located at $<25\,\%$ of the full absorption peak mean, can be explained by
    $\alpha$-particles stopped by the top ring of the source holder. This can
    happen in cases when a particle is emitted under a shallow angle relative
    to the source surface. As a consequence, they deposit only a fraction of
    their energy in the xenon gas, leading to a reduction of scintillation
    light. This hypothesis is qualitatively confirmed by using a Monte Carlo
    simulation to determine the track length distribution for particles which
    are stopped by the source holder and comparing it to the range in GXe
    (determined to be 21\,mm on average using SRIM~\cite{Ziegler:2010}),
    assuming the amount of light to be proportional to the track length.
    Event pile-up between these clipped tracks and full absorption events, as
    well as different light collection efficiencies depending on the 
    $\alpha$-particle direction, could be possible reasons for the asymmetric 
    shape of the full absorption peak.
    
    Besides the statistical errors from the fit, systematic errors affecting
    the average signal size are also taken into account. The most relevant one
    is due to variations in the distance between the $^{241\!}$Am source and
    the PTFE attenuator of up to 3\,mm in either direction. The related
    uncertainty was estimated using an optical simulation and amounts to 4\%.
    Another contribution (0.7\%) comes from an observed linear decrease in
    light yield during each measurement, likely caused by emanation of
    impurities into the xenon as soon as it is filled into the setup. Although
    the decrease is corrected for, the remaining error comes from possible
    variations of the measurement schedule which result in the time between
    data acquisition and setup filling to differ up to
    $\Delta t = 1 \text{min.}$ Furthermore, potential variation of the PMT's
    gain between measurements could introduce an additional uncertainty of
    0.6\%, as its operating voltage could only be set with an accuracy of
    $\Delta V = 1 \text{V}$. Finally, the impact of the fit range choice is
    also accounted for by varying the fit interval endpoints between
    $\pm 1\sigma \ldots \pm 3\sigma$ (up to 1.5\%). While the pressure of the
    filled xenon gas varied up to 0.6\%, the impact on the light yield is
    assumed to be negligible~\cite{Fernandes:2010}.

    \subsection{HeXe measurements}\label{subsec:hexe_data}
    To assess the PTFE transmittance in LXe, two sets of attenuators were
    inserted into the TPC described in Section~\ref{sec:setup}. Their
    respective thicknesses and positions within the detector are summarized in
    Table~\ref{tab:attenuators_hexe}. Given the inhomogeneous thickness
    introduced by a hole drilled into the 3.3\,mm thick disc covering the top
    PMT in the second measurement, this attenuator is not taken into account
    for further comparisons.

    \begin{table}[h]
    	\centering
    	\caption{PTFE attenuators used in the HeXe setup for comparison of their attenuation in LXe.}\label{tab:attenuators_hexe}
    	\begin{threeparttable}
    		\begin{tabular}{@{}p{\textwidth}@{}}
    			\centering
    			\begin{tabular}{ccllc}
    				attenuator	&	measurement				&	PMT		&	phase		&	thickness\\
    				\midrule
    				1			&	\multirow{2}{*}{a}	&	top		&	GXe			&	175\,$\mathrm{\upmu m}$\\
    				2			&			&	bottom	&	LXe			&	550\,$\mathrm{\upmu m}$\\
    				\hdashline
    				3			&	\multirow{2}{*}{b}	&	top		&	GXe			&	3.3\,mm\,\tnote{*}\\
    				4			&			&	bottom	&	LXe			&	700\,$\mathrm{\upmu m}$\\
    			\end{tabular}
    		\end{tabular}
    		\begin{tablenotes}
    			\centering
    			\item [*] Attenuator with inhomogeneous thickness due to a hole.
    		\end{tablenotes}
    	\end{threeparttable}
    \end{table}

\begin{figure}[ht]
	\begin{center}
		\includegraphics[angle=0,width=0.7\textwidth]{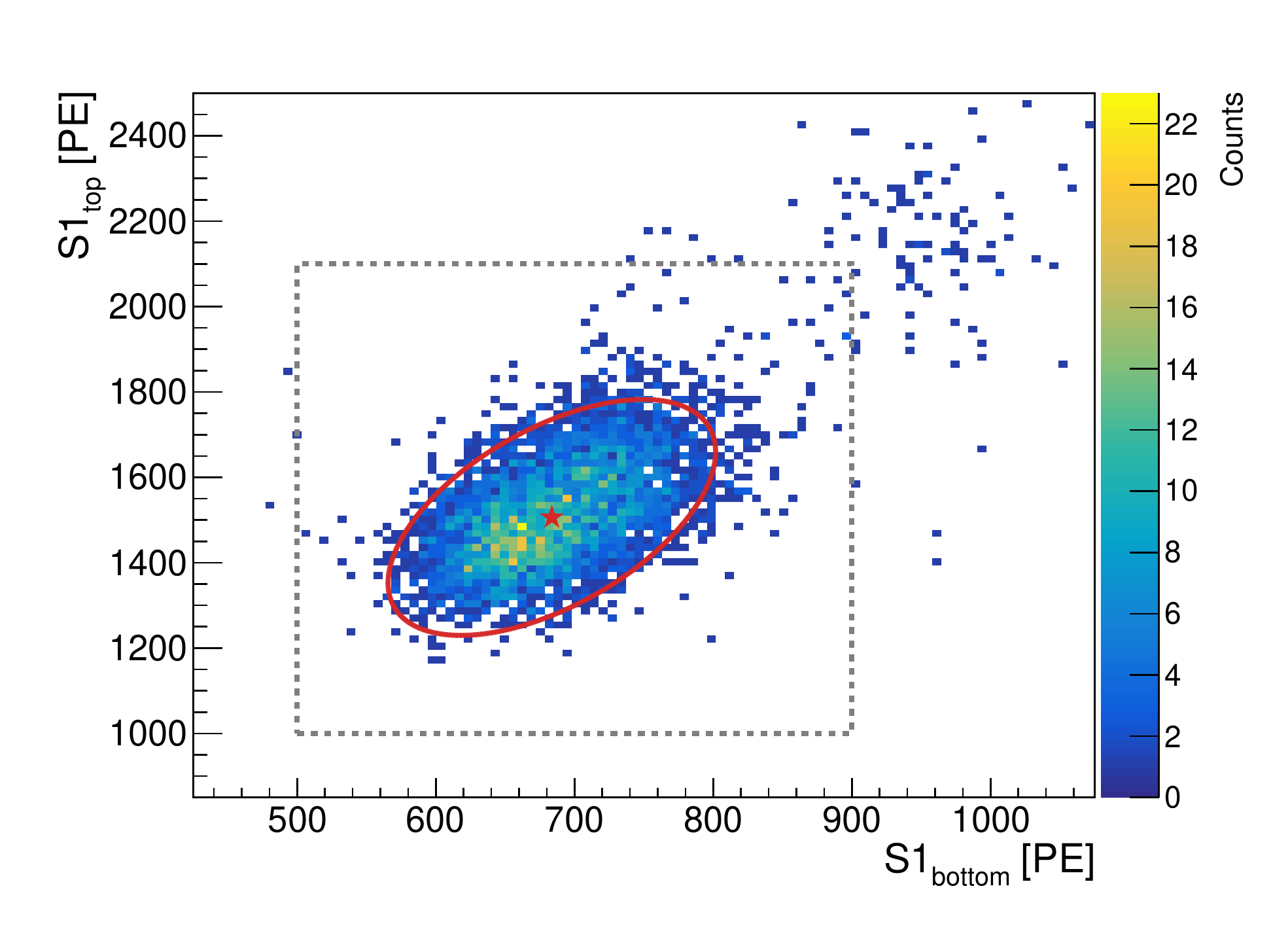}
		\caption[]{Detected signal size in the top and the bottom PMTs of the HeXe setup.
			$\alpha$-decays of $^{222}$Rn and its decay products happening in the central 2 mm of the TPC height are shown. A bivariate Gaussian function (red contour) is fitted to the data falling into the gray box. Its mean is highlighted by the red star. Data corresponds to \textit{measurement a} (see Table~\ref{tab:attenuators_hexe}).}\label{fig:hexe_plot_fit}
	\end{center}
	\end{figure}

    Energy depositions in the TPC's sensitive volume by $\alpha$-decays of $^{222}$Rn and its daughters $^{218}$Po and $^{214}$Po were recorded during dual-phase operation.
    The average amount of light seen by each photosensor
    is then determined by first selecting decays which happened within $\pm 1$\,
    mm around the central height of the TPC. Then, the amount of prompt scintillation
    light seen by the top PMT is plotted against the amount seen by the bottom
    one. Finally, a 2D Gaussian function is fitted to the population corresponding to
    the $\alpha$-decays of $^{222}$Rn and $^{218}$Po 
    (see Figure~\ref{fig:hexe_plot_fit}). These decays are grouped together as 
    their expected full absorption peaks cannot be resolved separately.
    
    To compare the average amount of light seen in HeXe with the room
    temperature setup data, the expected amount of light for $^{222}$Rn
    (daughter) $\alpha$-decays in HeXe without attenuators needs to be
    estimated. As  such measurements suffer from signal loss due to PMT
    saturation effects, this quantity can only be determined indirectly.
    Therefore, the signal size of $^{83m}$Kr decays as recorded by the
    top/bottom PMT in the HeXe setup without attenuators
    ($LY^{\text{HeXe}}(\text{Kr})_{t,b}$), is multiplied with the signal size
    ratio between $^{222}$Rn and $^{83m}$Kr signals as observed in
    XENON100\footnote{$LY^{\text{XE100}}(\text{Kr})/LY^{\text{XE100}}(\text{Rn}) = 173 \pm 4$}~\cite{Aprile:2017xxh,Aprile:2017fhu}.
    In the latter, PMT saturation effects when recording $\alpha$-decays were
    significantly less pronounced and could be corrected for. Decays of
    $^{83m}$Kr cannot be used to compare light levels with and without
    attenuator on their own, as they generate too little light which makes
    them non-detectable when the attenuators mentioned in
    Table~\ref{tab:attenuators_hexe} are present. In addition, the signal needs
    to be corrected for the modified probability for a photon to reach the top
    or the bottom PMT due to reflection off the PTFE attenuators. This is
    necessary to disentangle the effects of both reflection and absorption on
    the photon detection efficiency. The correction is applied by multiplying
    the amount of light seen by each PMT with a factor which depends on the
    probability for a photon to hit the PMT if no absorption were to take place
    in either attenuator (called $P$ from now on). An optical
    GEANT4~\cite{Agostinelli:2002hh} simulation
    of the HeXe setup was used to study how the presence of PTFE attenuators
    modifies $P$. Absorption inside the simulated PTFE attenuators was disabled.
    The correction factor is given as the ratio between
    $P(\text{no attenuators})$ and $P(\text{with attenuators})$. The
    attenuation factor $\gamma_i$ for the PTFE piece in front of
    PMT $i = \{t,b\}$ can then be calculated using the following equation:

    \begin{align}
        \gamma_{i} = \left[\frac{LY^{\text{HeXe}}(\text{Rn})}{LY^{\text{HeXe}}(\text{Kr})}\;\right]_{i = \{t,b\}} \cdot \left[\frac{P(\text{no attenuators})}{P(\text{with attenuators})}\;\right]_{i = \{t,b\}} \cdot \frac{LY^{\text{XE100}}(\text{Kr})}{LY^{\text{XE100}}(\text{Rn})}.\label{eq:hexe_scaling}
    \end{align}

    Because no dedicated reflectivity measurements of the PTFE pieces could be
    made, a range of potential values taken from attenuation measurement data
    and literature has to be taken into account. Reflectivity of PTFE in the
    GXe phase is assumed to be 76\% based on room temperature setup data
    (explained in Section~\ref{sec:results}). For PTFE without a dedicated
    surface finishing, a reflectivity of 96\% is assumed when immersed in LXe,
    similar to values reported for example by~\cite{Neves:2016tcw}. To quantify
    the uncertainty due to the range of potential reflectivities, the assumed
    values are varied by $\mathrm{\pm\,30\%}$, covering the range of typical
    literature values for the reflectivity of
    PTFE~\cite{Neves:2016tcw, Silva:2009ip}. This yields an estimate for the
    relative systematic uncertainty of about 50\%. The impact of variations in
    the liquid level height as well as the reflectivity of the stainless steel
    meshes was investigated as well, but found to be subdominant.

    \section{Results and discussion}\label{sec:results}
\begin{figure}[ht]
	\begin{center}
		\includegraphics[angle=0,width=0.7\textwidth]{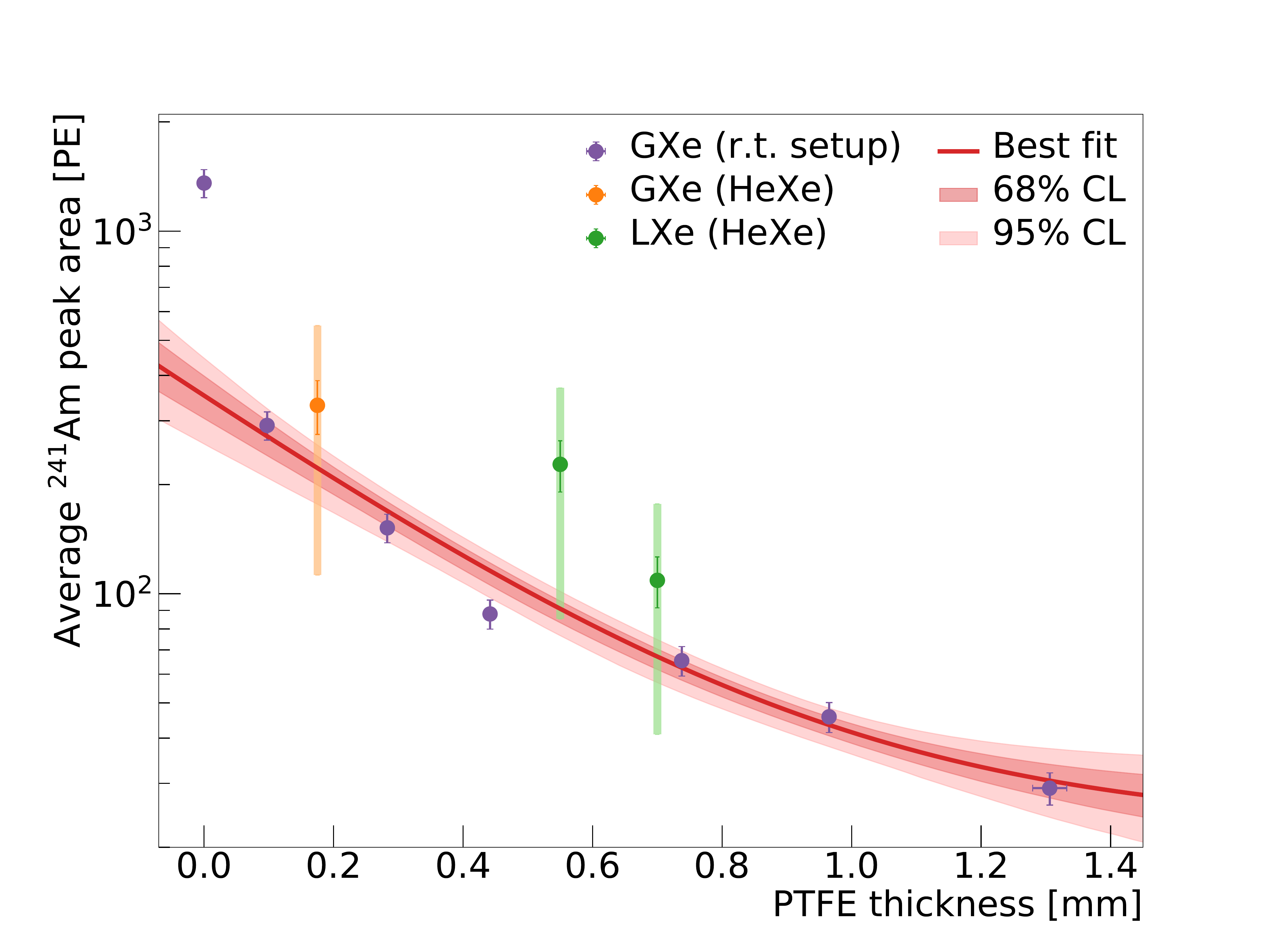}
		\caption[]{Average $\alpha$-decay xenon scintillation signal size
			in photo electrons (PE) versus PTFE attenuator
			thickness. The data acquired at room temperature
			(purple markers) are fitted with a modified Beer-Lambert
			function (equation~\ref{eq:lambert_law}), where the red
			shaded regions are the respective confidence bands of
			the fit function. For data points inferred using the
			HeXe setup, statistical uncertainties are shown
			separately from the combined error. The contribution of
			the former is indicated with vertical bars in the same
			color as the data point, while the combination of both
			the statistical uncertainty and systematic uncertainties
			from the optical simulation is shown using a brighter
			shade.}\label{fig:results}
	\end{center}
	\end{figure}

    The observed average $\alpha$-decay signal size is calculated as described
    in Sections~\ref{subsec:room_temp_setup_data} and~\ref{subsec:hexe_data} as a
    function of the thickness of the PTFE attenuator used for the corresponding
    measurement. Figure~\ref{fig:results} shows the data of the room
    temperature setup in purple, while the measurements in the HeXe setup are
    shown in orange and green for attenuators placed in the gas and liquid
    xenon phase, respectively.

    Assuming the Beer-Lambert law to be applicable, the amount of light $I$
    transmitted through a PTFE attenuator of thickness $d$ is modelled by:
    \begin{equation}
        I(d) = A \cdot e^{-\frac{d}{\lambda_t}} + B \label{eq:lambert_law},
    \end{equation}
    where $A$ denotes the amount of transmitted light expected for an
    infinitesimally thin attenuator and $\lambda_t$ represents the effective transmission
    coefficient of the medium. The constant offset $B$ allows for scattered
    light bypassing the attenuator to be taken into account which is found to
    be necessary to explain the data. Equation~\ref{eq:lambert_law} is fitted to
    the room temperature data (purple points, excluding the data point taken
    without attenuator) by minimizing $\upchi^2$ in order to extract the
    transmission coefficient, giving
    \begin{center}
        \resulta,
    \end{center}
    Here the statistical uncertainty of $\pm 50\,\mathrm{\upmu m}$ is determined
    from the fit, taking into account variations of the PTFE thickness, as
    well as the uncertainties on the light level which are detailed in Section~\ref{subsec:room_temp_setup_data}.
    The constant $B$ is found to be $(23\pm 6)\,\mathrm{PE}$.

    The data point obtained without attenuator deviates from the value expected by 
    extrapolation of the fit result, which is given by parameter $A$ in 
    Equation~\ref{eq:lambert_law} and amounts to $(330\pm 50)\,\mathrm{PE}$.
    We explain this deviation by the Fresnel reflection at the PTFE surface. 
    For all the measurements with an attenuator, part of the incoming scintillation 
    light is lost due to reflection. Using this measurement, a reflectivity 
    of about 76\% of the used PTFE in GXe is found.

    Since scintillation light is produced isotropically along the track length
    of the $\alpha$-particles, photons do not always arrive perpendicular with
    respect to the attenuator surface. If scattering of the light inside the PTFE
    is not taken into account, the amount of PTFE
    traversed before reaching the PMT depends on the angle of incidence,
    resulting in a larger effective thickness of the attenuators. The
    distribution of incident angles has been determined using an optical
    GEANT4 simulation of the room temperature setup, assuming isotropic
    emission of photons within a hemispheric region around the $^{241\!}$Am
    source having a radius of 21\,mm. The mean traversed PTFE thickness varies
    for different assumptions made on the reflective properties of the
    stainless steel tube. To get an estimate for the reflectivity of stainless
    steel for xenon scintillation light, the measurements reported
    in~\cite{Karlsson:1982rss} were extrapolated to the wavelength of 175\,nm.
    Given the unpolished surface of the tube's steel and that no dedicated
    surface finishing has been applied, 30\,\% diffuse reflectivity were
    assumed. Under this assumption, the average traversed length would then be
    increased by a factor of 1.16, leading to a 16\% larger transmission
    coefficient. For the highest reported VUV reflectivity of stainless steel
    (57\%)~\cite{Bricola:2007rss} and under the conservative assumption of pure
    specular reflection, which was found to yield more extreme incident angles
    than pure diffuse reflection, the effective thickness would be larger by a
    factor of 1.47 compared to the geometric thickness. Since the exact
    reflective properties of the used steel are unknown, this effect can not be
    corrected for. It is therefore reflected by the asymmetric systematic
    uncertainty of up to $\mathrm{+ 60\,\upmu m}$ on the result.

    As part of the light is reflected off the PTFE from layers deeper than
    the surface~\cite{Kravitz:2019zqv}, the total amount of reflected light
    would depend on the sample thickness. This would cause the transmittance
    to differ from the one derived from the Beer-Lambert law.
    Since the used setup does not allow to simultaneously measure the reflectance
    of the samples, this ambiguity can not be resolved.
    However, an estimation of the effect due to in-medium scattering on the 
    transmittance was done, using the so-called \textit{two-flux} model originally 
    proposed by Kubelka and Munk~\cite{Kubelka:1931,Kubelka:1948}. 
    This one dimensional model allows to predict the transmittance through 
    a parallel plane sample. Absorption within the medium is described by
    an absorption length $\lambda_a$, whereas scattering of the light into the 
    forward and backward direction is described by a total effective scattering
    length $\lambda_s$. The transmittances observed in the room temperature setup 
    are found to be well described by this model if a total effective scattering length of
    $\lambda_s \sim 50\,\mathrm{\upmu m}$ together with an absorption length 
    $\lambda_a \gg \lambda_s$ are assumed. Assuming this model is valid, we find
    that the predicted transmittance would be greater at larger thicknesses
    compared to the simple Beer-Lambert-based model which uses $\lambda_t$
    when disregarding the offset $B$. The evolution of the transmittance
    predicted by the two-flux model is compatible with the result reported
    in~\cite{Althueser:2020ovv}.

    Finally, in order to compare data obtained from both setups with each
    other, HeXe data points need to be scaled in order to match the light
    level of the room temperature setup. Consequently, the attenuation factor
    of each attenuator in the HeXe setup is first determined as described
    in Section~\ref{subsec:hexe_data}. It is then applied to the expected
    $^{241\!}$Am signal size in the room temperature setup without attenuator,
    estimated by extrapolation of the fit function to a PTFE thickness of
    0\,mm. All three data points agree with the fitted line within the combined
    systematic and $1\upsigma$ statistical error. It should be noted, that the 
    given thicknesses do not include the shrinkage of PTFE at cryogenic
    temperatures.

    \section{Conclusion}
    The transmission coefficient of PTFE for xenon scintillation light has been
    determined for the first time using a setup filled with
    GXe at room temperature, yielding:
    \begin{center}
        \resulta.
    \end{center}
    Complementary measurements with PTFE attenuators inside the HeXe LXe TPC
    were conducted in addition to verify the result under realistic detector
    conditions. Their results agree within the statistical and systematic
    errors with values expected when using the transmission coefficient
    estimate.

    While PTFE is a commonly-used material in LXe detectors due to its high
    reflectivity for xenon scintillation light, it is desired to reduce its
    amount to minimize radiogenic background. Knowledge of the transmission of
    xenon scintillation light thus helps in optimizing the design of future
    detectors such as DARWIN~\cite{Aalbers:2016jon}. The results obtained in
    this study imply, that PTFE of a few mm thickness is sufficient to
    optically decouple detector regions from each other.

    \acknowledgments{}
    We wish to acknowledge the support of the the Max Planck Society.
    We thank our technician Michael Rei\ss{}felder for his considerable support
    regarding both of the experimental setups as well as Lutz Alth\"user for
    his comments regarding the effective PTFE thickness.

    \bibliography{manuscript_v1.4}

\providecommand{\href}[2]{#2}\begingroup\raggedright\begin{thebibliography}{10}

\bibitem{Undagoitia:2015gya}
T.~Marrod\'an~Undagoitia and L.~Rauch, \emph{{Dark matter direct-detection
  experiments}},
  \href{http://dx.doi.org/10.1088/0954-3899/43/1/013001}{\emph{J. Phys.}
  {\bfseries G43} (2016) 013001},
  [\href{https://arxiv.org/abs/1509.08767}{{\ttfamily 1509.08767}}].

\bibitem{Pas:2015eia}
{H. P\"as, and W. Rodejohann}, \emph{{Neutrinoless Double Beta Decay}},
  \href{http://dx.doi.org/10.1088/1367-2630/17/11/115010}{\emph{New J. Phys.}
  {\bfseries 17} (2015) 115010},
  [\href{https://arxiv.org/abs/1507.00170}{{\ttfamily 1507.00170}}].

\bibitem{Aprile:2018dbl}
{\scshape XENON} collaboration, E.~Aprile et~al., \emph{{Dark Matter Search
  Results from a One Ton-Year Exposure of XENON1T}},
  \href{http://dx.doi.org/10.1103/PhysRevLett.121.111302}{\emph{Phys. Rev.
  Lett.} {\bfseries 121} (2018) 111302},
  [\href{https://arxiv.org/abs/1805.12562}{{\ttfamily 1805.12562}}].

\bibitem{Albert:2017owj}
{\scshape EXO} collaboration, J.~B. Albert et~al., \emph{{Search for
  Neutrinoless Double-Beta Decay with the Upgraded EXO-200 Detector}},
  \href{http://dx.doi.org/10.1103/PhysRevLett.120.072701}{\emph{Phys. Rev.
  Lett.} {\bfseries 120} (2018) 072701},
  [\href{https://arxiv.org/abs/1707.08707}{{\ttfamily 1707.08707}}].

\bibitem{Akerib:2017xx}
{\scshape LUX} collaboration, D.~S. Akerib et~al., \emph{{Results from a Search
  for Dark Matter in the Complete LUX Exposure}},
  \href{http://dx.doi.org/10.1103/PhysRevLett.118.021303}{\emph{Phys. Rev.
  Lett.} {\bfseries 118} (2017) 021303},
  [\href{https://arxiv.org/abs/1608.07648}{{\ttfamily 1608.07648}}].

\bibitem{Cui:2017xx}
{\scshape PandaX-II} collaboration, X.~Cui et~al., \emph{{Dark Matter Results
  from 54-Ton-Day Exposure of PandaX-II Experiment}},
  \href{http://dx.doi.org/10.1103/PhysRevLett.119.181302}{\emph{Phys. Rev.
  Lett.} {\bfseries 119} (2017) 181302},
  [\href{https://arxiv.org/abs/1708.06917}{{\ttfamily 1708.06917}}].

\bibitem{Fujii_2015xx}
F.~Keiko et~al., \emph{High-accuracy measurement of the emission spectrum of
  liquid xenon in the vacuum ultraviolet region},
  \href{http://dx.doi.org/10.1016/j.nima.2015.05.065}{\emph{Nucl. Instrum.
  Meth.} {\bfseries A795} (2015) 293}.

\bibitem{Jortner:1965xsw}
J.~Jortner et~al., \emph{{Localized Excitations in Condensed Ne, Ar, Kr, and
  Xe}}, \href{http://dx.doi.org/10.1063/1.1695927}{\emph{J. Chem. Phys.}
  {\bfseries 42} (1965) 4250--4253}.

\bibitem{Basov:1970}
N.~{Basov} et~al., \emph{{Luminescence of condensed Xe, Kr, Ar and their
  mixtures in vacuum region of spectrum under excitation by fast electrons}},
  \href{http://dx.doi.org/10.1016/0022-2313(70)90095-5}{\emph{J. Lumin.}
  {\bfseries 1-2} (1970) 834--841}.

\bibitem{Aprile:2017aty}
{\scshape XENON} collaboration, E.~Aprile et~al., \emph{{The XENON1T Dark
  Matter Experiment}},
  \href{http://dx.doi.org/10.1140/epjc/s10052-017-5326-3}{\emph{Eur. Phys. J.}
  {\bfseries C77} (2017) 881},
  [\href{https://arxiv.org/abs/1708.07051}{{\ttfamily 1708.07051}}].

\bibitem{Akerib:2019fml}
{\scshape LZ} collaboration, D.~S. Akerib et~al., \emph{{The LUX-ZEPLIN (LZ)
  Experiment}}, \href{http://dx.doi.org/10.1016/j.nima.2019.163047}{\emph{Nucl.
  Instrum. Methods Phys. Res. A} {\bfseries 953} (2020) 163047},
  [\href{https://arxiv.org/abs/1910.09124}{{\ttfamily 1910.09124}}].

\bibitem{Cao:2014jsa}
{\scshape PandaX} collaboration, X.~Cao et~al., \emph{{PandaX: A Liquid Xenon
  Dark Matter Experiment at CJPL}},
  \href{http://dx.doi.org/10.1007/s11433-014-5521-2}{\emph{Sci. China Phys.
  Mech. Astron.} {\bfseries 57} (2014) 1476},
  [\href{https://arxiv.org/abs/1405.2882}{{\ttfamily 1405.2882}}].

\bibitem{Auger:2012gs}
M.~Auger et~al., \emph{{The EXO-200 detector, part I: Detector design and
  construction}},
  \href{http://dx.doi.org/10.1088/1748-0221/7/05/P05010}{\emph{JINST}
  {\bfseries 7} (2012) P05010},
  [\href{https://arxiv.org/abs/1202.2192}{{\ttfamily 1202.2192}}].

\bibitem{Silva:2009ip}
C.~Silva et~al., \emph{{Reflectance of Polytetrafluoroethylene (PTFE) for Xenon
  Scintillation Light}}, \href{http://dx.doi.org/10.1063/1.3318681}{\emph{J.
  Appl. Phys.} {\bfseries 107} (2010) 064902},
  [\href{https://arxiv.org/abs/0910.1056}{{\ttfamily 0910.1056}}].

\bibitem{Levy:2014}
C.~Levy, \emph{{Light propagation and reflection off Teflon in liquid xenon
  detectors for the XENON100 and XENON1T experiments}}, {\emph{PhD Thesis,
  University of M\"unster (Germany)} (2014) }.

\bibitem{Akerib:2012ak}
{\scshape LUX} collaboration, D.~S. Akerib et~al., \emph{{Technical Results
  from the Surface Run of the LUX Dark Matter Experiment}},
  \href{http://dx.doi.org/10.1016/j.astropartphys.2013.02.001}{\emph{Astropart.
  Phys.} {\bfseries 45} (2013) 34},
  [\href{https://arxiv.org/abs/1210.4569}{{\ttfamily 1210.4569}}].

\bibitem{Neves:2016tcw}
F.~Neves et~al., \emph{{Measurement of the absolute reflectance of
  polytetrafluoroethylene (PTFE) immersed in liquid xenon}},
  \href{http://dx.doi.org/10.1088/1748-0221/12/01/P01017}{\emph{JINST}
  {\bfseries 12} (2017) P01017},
  [\href{https://arxiv.org/abs/1612.07965}{{\ttfamily 1612.07965}}].

\bibitem{Aprile:2017gs}
{\scshape XENON} collaboration, E.~Aprile et~al., \emph{{Material Radioassay
  and Selection for the XENON1T Dark Matter Experiment}},
  \href{http://dx.doi.org/10.1140/epjc/s10052-017-5329-0}{\emph{Eur. Phys. J.
  C} {\bfseries 77} (2017) 890},
  [\href{https://arxiv.org/abs/1705.01828}{{\ttfamily 1705.01828}}].

\bibitem{Heaton:1989:an}
R.~Heaton et~al., \emph{{Neutron production from thick-target
  ({$\mathrm{\alpha}$}, n) reactions}},
  \href{http://dx.doi.org/https://doi.org/10.1016/0168-9002(89)90579-2}{\emph{Nucl.
  Instrum. Methods Phys. Res. A} {\bfseries 276} (1989) 529 -- 538}.

\bibitem{Agostini:2020adk}
{\scshape DARWIN} collaboration, F.~Agostini et~al., \emph{{Sensitivity of the
  DARWIN observatory to the neutrinoless double beta decay of $^{136}$Xe}},
  [\href{https://arxiv.org/abs/2003.13407}{{\ttfamily 2003.13407}}].

\bibitem{Aprile:2011dd}
{\scshape XENON} collaboration, E.~Aprile et~al., \emph{{The XENON100 Dark
  Matter Experiment}},
  \href{http://dx.doi.org/10.1016/j.astropartphys.2012.01.003}{\emph{Astropart.
  Phys.} {\bfseries 35} (2012) 573},
  [\href{https://arxiv.org/abs/1107.2155}{{\ttfamily 1107.2155}}].

\bibitem{Barrow:2016doe}
P.~Barrow et~al., \emph{{Qualification Tests of the R11410-21 Photomultiplier
  Tubes for the XENON1T Detector}},
  \href{http://dx.doi.org/10.1088/1748-0221/12/01/P01024}{\emph{JINST}
  {\bfseries 12} (2017) P01024},
  [\href{https://arxiv.org/abs/1609.01654}{{\ttfamily 1609.01654}}].

\bibitem{Auer:2020}
{Auer Kunststofftechnik GmbH \& Co. KG}, ``{Material datasheet PTFE}.''
  {http://www.auer-kunststofftechnik.de/pdf/Datenblatt PTFE.PDF}, {accessed
  2020-07-16}.

\bibitem{Ozone:2005ysa}
K.~Ozone, \emph{{Liquid Xenon Scintillation Detector for the New $\mu \to
  e\gamma$ Search Experiment}}.
\newblock PhD thesis, Tokyo U., 2005.

\bibitem{Cichon:2015}
D.~Cichon, \emph{Identifying $^{222}${R}n decay chain events in liquid xenon
  detectors},  Master's thesis, University of Heidelberg, 2015.

\bibitem{Joerg:2017}
F.~J\"org, \emph{Investigation of coating-based radon barriers and studies
  towards their applicability in liquid xenon detectors},  Master's thesis,
  University of Heidelberg, 2017.

\bibitem{Saldanha:2016mkn}
R.~Saldanha, L.~Grandi, Y.~Guardincerri and T.~Wester, \emph{{Model Independent
  Approach to the Single Photoelectron Calibration of Photomultiplier Tubes}},
  \href{http://dx.doi.org/10.1016/j.nima.2017.02.086}{\emph{Nucl. Instrum.
  Meth.} {\bfseries A863} (2017) 35},
  [\href{https://arxiv.org/abs/1602.03150}{{\ttfamily 1602.03150}}].

\bibitem{Manalaysay:2009yq}
A.~Manalaysay et~al., \emph{{Spatially uniform calibration of a liquid xenon
  detector at low energies using 83m-Kr}},
  \href{http://dx.doi.org/10.1063/1.3436636}{\emph{Rev. Sci. Instrum.}
  {\bfseries 81} (2010) 073303},
  [\href{https://arxiv.org/abs/0908.0616}{{\ttfamily 0908.0616}}].

\bibitem{Kastens:2009rt}
L.~W. Kastens, S.~Bedikian, S.~B. Cahn, A.~Manzur and D.~N. McKinsey, \emph{{A
  83Krm Source for Use in Low-background Liquid Xenon Time Projection
  Chambers}},
  \href{http://dx.doi.org/10.1088/1748-0221/5/05/P05006}{\emph{JINST}
  {\bfseries 5} (2010) P05006},
  [\href{https://arxiv.org/abs/0912.2337}{{\ttfamily 0912.2337}}].

\bibitem{Bruenner:2020}
S.~Bruenner et~al., \emph{{Radon daughter removal from PTFE surfaces and its
  application in liquid xenon detectors}}, {\emph{in preparation} (expected
  2020) }.

\bibitem{Ziegler:2010}
J.~F. {Ziegler}, M.~D. {Ziegler} and J.~P. {Biersack}, \emph{{SRIM - The
  stopping and range of ions in matter (2010)}},
  \href{http://dx.doi.org/10.1016/j.nimb.2010.02.091}{\emph{Nucl. Instrum.
  Meth.} {\bfseries B268} (2010) 1818}.

\bibitem{Fernandes:2010}
L.~M.~P. Fernandes et~al., \emph{Primary and secondary scintillation
  measurements in a xenon gas proportional scintillation counter},
  \href{http://dx.doi.org/10.1088/1748-0221/5/09/p09006}{\emph{JINST}
  {\bfseries 5} (2010) P09006}.

\bibitem{Aprile:2017xxh}
{\scshape XENON} collaboration, E.~Aprile et~al., \emph{{Signal Yields of keV
  Electronic Recoils and Their Discrimination from Nuclear Recoils in Liquid
  Xenon}}, \href{http://dx.doi.org/10.1103/PhysRevD.97.092007}{\emph{Phys.
  Rev.} {\bfseries D97} (2018) 092007},
  [\href{https://arxiv.org/abs/1709.10149}{{\ttfamily 1709.10149}}].

\bibitem{Aprile:2017fhu}
{\scshape XENON} collaboration, E.~Aprile et~al., \emph{{Intrinsic backgrounds
  from Rn and Kr in the XENON100 experiment}},
  \href{http://dx.doi.org/10.1140/epjc/s10052-018-5565-y}{\emph{Eur. Phys. J.
  C} {\bfseries C78} (2018) 132},
  [\href{https://arxiv.org/abs/1708.03617}{{\ttfamily 1708.03617}}].

\bibitem{Agostinelli:2002hh}
{\scshape GEANT4} collaboration, S.~Agostinelli et~al., \emph{{GEANT4: A
  Simulation toolkit}},
  \href{http://dx.doi.org/10.1016/S0168-9002(03)01368-8}{\emph{Nucl. Instrum.
  Meth. A} {\bfseries 506} (2003) 250--303}.

\bibitem{Karlsson:1982rss}
B.~Karlsson and C.~G. Ribbing, \emph{Optical constants and spectral selectivity
  of stainless steel and its oxides},
  \href{http://dx.doi.org/10.1063/1.331503}{\emph{J. Appl. Phys.} {\bfseries
  53} (1982) 6340--6346}.

\bibitem{Bricola:2007rss}
S.~Bricola et~al., \emph{{Noble-gas liquid detectors: measurement of light
  diffusion and reflectivity on commonly adopted inner surface materials}},
  \href{http://dx.doi.org/https://doi.org/10.1016/j.nuclphysbps.2007.08.059}{\emph{Nucl.
  Phys. B Proc. Suppl.} {\bfseries 172} (2007) 260 -- 262}.

\bibitem{Kravitz:2019zqv}
S.~Kravitz, R.~Smith, L.~Hagaman, E.~Bernard, D.~McKinsey, L.~Rudd et~al.,
  \emph{{Measurements of Angle-Resolved Reflectivity of PTFE in Liquid Xenon
  with IBEX}},
  \href{http://dx.doi.org/10.1140/epjc/s10052-020-7800-6}{\emph{Eur. Phys. J.
  C} {\bfseries 80} (2020) 262},
  [\href{https://arxiv.org/abs/1909.08730}{{\ttfamily 1909.08730}}].

\bibitem{Kubelka:1931}
P.~Kubelka and F.~Munk, \emph{{Ein Beitrag zur Optik der Farbanstriche}},
  {\emph{{Zeitschrift f\"ur technische Physik}} {\bfseries 12} (1931)
  593--601}.

\bibitem{Kubelka:1948}
P.~Kubelka, \emph{{New Contributions to the Optics of Intensely
  Light-Scattering Materials. Part I}},
  \href{http://dx.doi.org/10.1364/JOSA.38.000448}{\emph{J. Opt. Soc. Am.}
  {\bfseries 38} (May, 1948) 448--457}.

\bibitem{Althueser:2020ovv}
L.~Althueser, S.~Lindemann, M.~Murra, M.~Schumann, C.~Wittweg and
  C.~Weinheimer, \emph{{VUV Transmission of PTFE for Xenon-based Particle
  Detectors}},  [\href{https://arxiv.org/abs/2006.05827v1}{{\ttfamily
  2006.05827v1}}].

\bibitem{Aalbers:2016jon}
{\scshape DARWIN} collaboration, J.~Aalbers et~al., \emph{{DARWIN: Towards the
  Ultimate Dark Matter Detector}},
  \href{http://dx.doi.org/10.1088/1475-7516/2016/11/017}{\emph{JCAP} {\bfseries
  1611} (2016) 017}, [\href{https://arxiv.org/abs/1606.07001}{{\ttfamily
  1606.07001}}].

\end{thebibliography}\endgroup
    \bibliographystyle{JHEP_mod}
\end{document}